\documentclass[a4paper,pra,showpacs,superscriptaddress,twocolumn]{revtex4-2}
\usepackage[utf8]{inputenc}
\usepackage[T1]{fontenc}

\usepackage{hyperref}
\usepackage{amsfonts,dsfont,mathrsfs,amsmath,amssymb,bm,bbm}
\usepackage[english]{babel}
\usepackage[babel]{microtype}  
\usepackage{graphicx}
\usepackage{subfigure}
\usepackage{braket}

\newcommand{\myref}[1]{Eq.\hspace{0.3em}\eqref{#1}}
\newcommand{\mycite}[1]{Ref.\hspace{0.3em}\cite{#1}}
\newcommand{\mycites}[1]{Refs.\hspace{0.3em}\cite{#1}}
\newcommand{\myfig}[1]{Fig.\hspace{0.3em}\ref{#1}}
\newcommand{\mysec}[1]{Sec.\hspace{0.3em}\ref{#1}}

\newcommand{\mc}[1]{\mathcal{#1}}

\begin{document}

	\title{Quantum complementarity from  a measurement-based perspective}
	
	\author{Shan Huang}
	\affiliation{National Laboratory of Solid State Microstructures and School of Physics,  Collaborative Innovation Center of Advanced Microstructures, Nanjing University, Nanjing 210093, China}
	\affiliation{Institute for Brain Sciences and Kuang Yaming Honors School, Nanjing University, Nanjing 210023,
		China}
    \affiliation{Hefei National Laboratory, University of Science and Technology of China, Hefei 230088, China}
    
	\author{Wen-Bo Liu}
	\affiliation{National Laboratory of Solid State Microstructures and School of Physics, Collaborative Innovation Center of Advanced Microstructures, Nanjing University, Nanjing 210093, China}
	
		\author{Yundu Zhao}
	\affiliation{National Laboratory of Solid State Microstructures and School of Physics, Collaborative Innovation Center of Advanced Microstructures, Nanjing University, Nanjing 210093, China}
	\affiliation{Institute for Brain Sciences and Kuang Yaming Honors School, Nanjing University, Nanjing 210023,
		China}
	
	\author{Hua-Lei Yin}
	
	\email{hlyin@nju.edu.cn}
	\affiliation{National Laboratory of Solid State Microstructures and School of Physics, Collaborative Innovation Center of Advanced Microstructures, Nanjing University, Nanjing 210093, China}
	
	\author{Zeng-Bing Chen}
	\email{zbchen@nju.edu.cn}
	\affiliation{National Laboratory of Solid State Microstructures and School of Physics, Collaborative Innovation Center of Advanced Microstructures, Nanjing University, Nanjing 210093, China}
	
	\author{Shengjun Wu}
	\email{sjwu@nju.edu.cn}
	\affiliation{National Laboratory of Solid State Microstructures and School of Physics, Collaborative Innovation Center of Advanced Microstructures, Nanjing University, Nanjing 210093, China}
	\affiliation{Institute for Brain Sciences and Kuang Yaming Honors School, Nanjing University, Nanjing 210023,
		China}
    \affiliation{Hefei National Laboratory, University of Science and Technology of China, Hefei 230088, China}

	\begin{abstract}
One of the most remarkable features of quantum physics is that attributes of quantum objects, such as the wave-like and particle-like behaviors of single photons, can be complementary in the sense that they are equally real but cannot be observed simultaneously. Quantum measurements, serving as windows providing views into the abstract edifice of quantum theory, are basic tools for manifesting the intrinsic behaviors of quantum objects. 
However, quantitative formulation of complementarity that highlights its manifestations in sophisticated measurements remains elusive.
Here we develop a general framework for demonstrating quantum complementarity in the form of information exclusion relations (IERs), which incorporates the wave-particle duality relations as particular examples. Moreover, we explore the applications of our theory in entanglement witnessing and elucidate that our IERs lead to an extended form of entropic uncertainty relations, providing intriguing insights into the connection between quantum complementarity and the preparation uncertainty.
	\end{abstract}
	
	\maketitle
	
  \section{Introduction}
	Quantum mechanics imposes fundamental limits on an observer's information gain in complementary measurements. In the light of  Bohr's complementarity principle \cite{bohr1928},  quantum systems possess mutually exclusive properties that are equally real, and a measurement to reveal one property would inevitably preclude all the complementary ones. Characterizing this subtle relationship between measurement strategy and information gain is significant for the sophisticated manipulation of quantum measurements in various tasks, from demonstrating genuine nonclassical features of quantum objects to general quantum information processing.

	Wootters and Zurek \cite{wootters1979} proposed the first quantitative statement of complementarity relation by taking an information-theoretical perspective into the competitive tradeoff between the wave-like and particle-like behaviors of  single photons. This  kind of wave-particle duality relations (WPDRs) are currently expressed in a concise inequality form \cite{greenberger1988,jaeger1993,jaeger1995,englert1996} for photons within the Mach-Zehnder interferometer (MZI; see \myfig{mzi}). For example, Jaeger \emph{et al} \cite{jaeger1993} and Englert \cite{englert1996} obtained the duality relation $\mc{V}^2+\mc{D}^2\leq1$ between fringe visibility (wave property) $\mc{V}$ and path distinguishability (particle property) $\mc{D}$. It is thus obvious that better which-way information implies less wave information, and vice versa.

	Heisenberg's uncertainty principle \cite{heisenberg1927} is another fundamental concept in quantum mechanics which captures similar underlying physics of complementarity. It states that outcomes of specific measurements, e.g., position and momentum of a single particle, cannot be predicted with certainty simultaneously. Modern formulations of the uncertainty principle typically use entropic uncertainty measures \cite{deutsch1983,maassen1988} due to their operational significance and the widespread applications \cite{coles2017} of entropic uncertainty relations (EURs), e.g., in the security analysis of quantum protocols \cite{cerf2002,berta2010,tomamichel2011}.

   The connections and contrasts between uncertainty and complementarity have been intensively debated \cite{storey1994,storey1995,englert1995,wiseman1995,busch2006,liu2012,coles2014}. It has been wondered whether novel complementarity relations can be derived directly from the well-studied and already-proven EURs. Particularly,  Coles \emph{et al} \cite{coles2014} proved that several WPDRs in the two-way interferometer can be equivalently reformulated as EURs for complementary observables. Thus two fundamental concepts of quantum mechanics are unified in this simple case.

	Nevertheless, entropy is a natural measure of lack of information regarding only observation-independent properties and becomes conceptually inadequate \cite{brukner2001} for quantum properties which are contextual and do not exist prior to measurements \cite{kochen1967, mermin1990}.
    To avoid this dilemma, Brukner and Zeilinger proposed an operationally invariant information measure of quantum systems \cite{brukner1999}. It is naturally aligned with the concept of complementarity as being elegantly defined as the sum of individual measures of information gain over a complete set of mutually unbiased bases (CMUBs) \cite{ivonovic1981,wootters1989,klappenecker2004,pittenger2004}---complementary bases---independent of particular choices of CMUBs and invariant under unitary time evolution. These intriguing properties inspired a series of insightful investigations  \cite{vrehavcek2002,lee2003,lee2000,giovannetti2004measurements,kofler2010,wang2019,ding2020,brukner2005}, including
   quantum state estimation \cite{vrehavcek2002,lee2003} and uncertainty relations for MUBs \cite{wang2019,ding2020}.

	In this paper, we adopt the operationally invariant information measure \cite{brukner1999} and develop a general framework for characterizing quantum complementarity beyond WPDRs, in terms of basic limits on one's ability to gain information about quantum systems through complementary measurement setups, i.e.,  information exclusion relations (IERs).  We emphasize that when considering generalized measurements, identifying certainty of outcome statistics with information gain or visibility of physical property faces conceptual challenge---an outcome predictable with 100\% certainty not necessarily reflects the complete information of the measured system.  In contrast to IERs  \cite{hall1995,hall1997,coles2014exclusion,berta2014,zhang2015} that utilize entropic mutual information or deriving complementarity relations from EURs \cite{coles2014}, our theory applies to generalized measurements and well captures the complete information of quantum systems as conserved quantities comprised of complementary pieces, highlighting the interplay between different pieces of information and their complementary nature.

	This paper is structured as follows. In  \mysec{pre}, we introduce some preliminary notations. In \mysec{povminfgain}, we propose a measure of information gain in individual measurements while formalizing the concept of complementary information. In   \mysec{secier}, we proceed to establish IERs which restrict one's weighted sum of information gains over multiple measurements, with and without quantum memory respectively.  In  \mysec{originWPDRs}, we show how our IERs lead to tight WPDRs. In \mysec{discussions}, we discuss practical applications of our IERs. Finally, we briefly conclude this work in \mysec{conclusion}.

  \section{Preliminary}\label{pre}

   On a $d$-dimensional Hilbert space $\mc{H}_d$, each generalized measurement, i.e., positive-operator-valued measure (POVM), is a collection of positive semi-definite operators (called effects) $\mc{M}=\{M_i\}$ that sum up to the identity operator:  $M_i\geq0$ and $\sum_iM_i=\mathbbm{1}_d$. In particular, the measurement of a nondegenerate observable is described by rank-1 projectors onto its eigenvectors, i.e., rank-1 projective measurement.  When a quantum state $\rho$ is measured, the outcome probabilities are given by Born's rule, $p_i=tr(M_i\rho)$.

	The Choi-Jamiołkowski isomorphism  \cite{choi1972} allows us to represent each operator $O$   on $\mc{H}_d$ as a vector $\ket{O}$ of the product space $\mc{H}_d^{\otimes2}$:
	\begin{align}
		\ket{O}&=\sqrt{d}\cdot O\otimes\mathbbm{1}_d\ket{\psi_d}=\sum_{i,j=1}^{d}O_{i,j}\ket{i}\otimes\ket{j}^*,  \label{isom}\\
		 O&=\sqrt{d}\cdot tr_2\left(|O\rangle\langle\psi_d|\right),  \nonumber
	\end{align}
	where $\ket{\psi_d}=\frac{1}{\sqrt{d}}\sum_{i=1}^{d}\ket{i}\otimes\ket{i}^*$  is the maximally entangled isotropic  state and $tr_2(\cdot)$ denotes the partial trace over the second space.  A useful property of \myref{isom} that will be exploited is that  $\braket{O_1|O_2}=tr(O_1^+O_2)$ holds for any two operators $O_1$ and $O_2$ on $\mc{H}_d$.\\

   \section{Measure of information gain}\label{povminfgain}

   In the light of Kochen-Specker's theorem \cite{kochen1967} (see also \mycite{mermin1990}), it is impossible to assign a definite noncontextual value to every quantum observable. During a measurement, all that an observer has is the probabilistic occurrence of one outcome (labeled contextual value 1), which simultaneously negates the occurrence of other outcomes (labeled contextual values 0).  The information content of quantum systems is thus reflected in the  statistics of these contextual binary strings.

  Consider an experimental setup to perform the measurement $\mc{M}=\{M_i\}$ on individual copies of a quantum state  that is unknown to the experimenter.  Each time the $i$th outcome occurs, the experimenter gets a squared deviation $(1-tr(M_i)/d)^2$ from the expectation $tr(M_i)/d$ for the completely mixed state---least information state---or gets $(0-tr(M_i)/d)^2$ otherwise. After repeating the experiments large enough $N$ times, the total squared deviation is  $D_i^2=N[p_i(1-tr(M_i)/d)^2+(1-p_i)(tr(M_i)/d)^2]$, which consists of two contributions  $D_i^2=\Delta_i^2+B_i^2$. Wherein $\Delta_i^2=N[p_i(1-p_i)]$ is the total uncertainty (variance),  which determines the width $2\Delta_i/N$ of the confidence interval $[p_i-\frac{1}{N}\Delta_i,p_i+\frac{1}{N}\Delta_i]$ for estimating the outcome probabilities $\{p_i\}$.

	What truly discriminates the measured state from the completely mixed state, on the other hand, is the total squared bias $B_i^2= N(p_i-tr(M_i)/d)^2$.  We suggest the measure of information gain on the state $\rho$  in each individual trial of the measurement  $\mc{M}=\{M_i\}$ to be the sum of mean squared bias over all outcomes
     \begin{align}
		G(\mc{M})_\rho&=\sum_{i}\big( p_i-tr(M_i)/d\big)^{2}=: \braket{\rho|\hat{G}(\mc{M})|\rho}. \label{infgain}
	\end{align}
In the above,  we leverage the isomorphism \eqref{isom} to define the view operator of a measurement $\mc{M}$ as 
	\begin{align}
		\hat{G}(\mc{M})=\sum_i|\tilde{M}_i\rangle\langle \tilde{M}_i|
		\label{viewoperator}
	\end{align}
 where $\tilde{M}_{i}=M_{i}-\frac{1}{d}tr(M_i)\mathbbm{1}_d$ is traceless, or equivalently, $\ket{\tilde{M}_{i}}=\ket{M_{i}}-|\psi_d\rangle\langle\psi_d|M_{i}\rangle$ is orthogonal to $\ket{\psi_d}$. View operators are positive semi-definite,  $\hat{G}\geq0$ on the $(d^2-1)$-dimensional subspace $\mc{H}_{\perp\psi_d}$ of $\mc{H}_d^{\otimes2}$ orthogonal to $\ket{\psi_d}$, and vanish for trivial POVMs whose effects are all proportional to the identity, $M_i=\frac{1}{d}tr(M_i)\mathbbm{1}_d$.

\begin{figure}[t]
	\includegraphics[width=8.6cm]{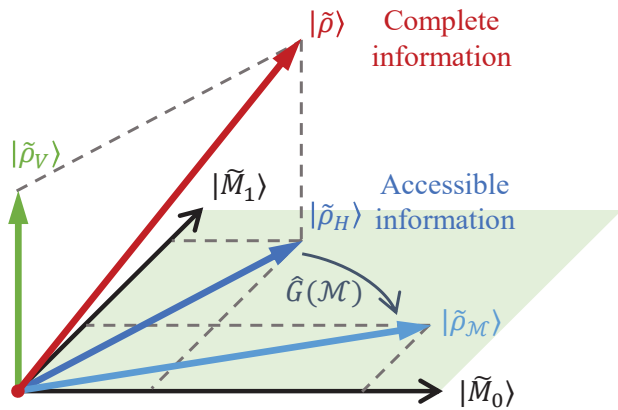}
	\caption{Illustration of the information complementarity, where the vector $\ket{\tilde{\rho}}$ encodes the complete information content of the state $\rho$.  For a two-outcome measurement $\mc{M}=\{M_i\}_{i=0}^1$, the vectors $\{\ket{\tilde{M}_i}\}$ span a 2-dimensional space (colored horizontal plane), on which the view operator $\hat{G}(\mc{M})$ is a bijective transform. While the horizontal component  $\ket{\tilde{\rho}_H}$ of $\ket{\tilde{\rho}}$ can be reconstructed from the vector $\ket{\tilde{\rho}_{\mc{M}}}$ encoding  the outcome statistics, its vertical component $\ket{\tilde{\rho}_V}$ contains only information complementary to what is accessible through $\mc{M}$.}
	\label{viewvector}
\end{figure}

Now we are able to formalize our idea of complementary information.  Let $\tilde{\rho}=\rho-\mathbbm{1}_d/d$, observe that the outcome probabilities of a measurement $\mathcal{M}$ on the state $\rho$ are encoded in the expansion coefficients of the vector $\ket{\tilde{\rho}_{\mathcal{M}}} = \hat{G}(\mathcal{M}) \cdot \ket{\rho} = \hat{G}(\mathcal{M}) \cdot \ket{\tilde{\rho}} = \sum_{i}(p_i - \text{tr}(M_i)/d) \ket{\tilde{M}_i}$ under the basis $\{\ket{\tilde{M}_i}\}$. The vector $\ket{\tilde{\rho}_{\mathcal{M}}}$ encodes the complete information of $\rho$ if $\ket{\tilde{\rho}}$ lies in the subspace of $\mc{H}_{\perp\psi_d}$ on which the view operator $\hat{G}(\mathcal{M})$ is invertible, whereas if $\ket{\tilde{\rho}}$ is orthogonal to that space,  $\ket{\tilde{\rho}_{\mathcal{M}}}$ vanishes and $\mathcal{M}$ cannot be employed to distinguish $\rho$ from the completely mixed state (see \myfig{viewvector} for an illustration of the geometric relations between the above vectors). In the sense above, two nontrivial measurements $\mc{M}_1$ and $\mc{M}_2$ satisfying
\begin{align}
   \hat{G}(\mc{M}_1)\cdot\hat{G}(\mc{M}_2)=0
   \label{complementary}
 \end{align}
are complementary since, if the complete information of  $\rho$ is accessible through $\mc{M}_1$, then no information gain is accessible through $\mc{M}_2$, and vice versa. We prove in Appendix. \ref{appview} that measurements in mutually unbiased bases (MUBs) \cite{ivonovic1981,wootters1989,klappenecker2004,pittenger2004} are complementary.

It is worth mentioning that the combined view operator $\hat{G}=\sum_{\theta}\hat{G}(\mc{M}_\theta)$ associated with a set of  POVMs  $\mathscr{M}=\{M_{\theta}\}$ on $\mc{H}_d$ can be positive definite (invertible) on  $\mc{H}_{\perp\psi_d}$. In this case, no POVM can be complementary to all POVMs of $\mathscr{M}$ simultaneously. This means that $\mathscr{M}$ is informationally-complete and $\hat{G}$ offers a complete view to all $d$-dimensional quantum states. Utilizing the isomorphism \eqref{isom}, arbitrary unknown state $\rho$ can then be reconstructed from the vector $\hat{G}\ket{\tilde{\rho}}=\ket{\tilde{\rho}_{\mathscr{M}}}$ encoding the outcome statistics as follows
\begin{align}
\rho=\sqrt{d}\cdot tr_2(\hat{G}^{-1}|\tilde{\rho}_{\mathscr{M}}\rangle\langle\psi_d|)+\mathbbm{1}_d/d.          \label{tomo}
\end{align}
For further readings on the topic of state estimation, we recommend \mycites{Ariano2001, Ariano2004}.

 Interestingly, the combined view operator associated with CMUBs of $\mc{H}_d$, i.e., $d+1$ mutually unbiased bases (MUBs) \cite{ivonovic1981,wootters1989,klappenecker2004,pittenger2004}, is simply the identity operator $\mathbbm{1}_{\perp\psi_d}=\mathbbm{1}_d\otimes \mathbbm{1}_d-|\psi_d\rangle\langle\psi_d|$ on $\mc{H}_{\perp\psi_d}$ (see Appendix. \ref{appview}). Thus the operationally invariant measure \cite{brukner1999} of complete information content contained in quantum states can be restateted in our language as
	\begin{align}
			I_{\rm com}(\rho)=\braket{\rho|\mathbbm{1}_{\perp\psi_d}|\rho}=tr(\rho^2)-1/d. \label{cominf}
  \end{align}
This measure naturally coincides with Bohr's idea \cite{bohr1928} that only the totality of complementary properties together exhausts the complete information of objects.

    \section{Information exclusion relations} \label{secier}
	To formulate quantum complementarity into information exclusion relations, next we focus on the measurement scenarios where distinct measurements on individual copies of a quantum system are selected with biased (non-uniform) probabilities.
	
	\subsection{Local information exclusion relations}

	\emph{Theorem 1.} For a set of measurements $\{\mc{M}_{\theta}\}$ with selection probabilities $\{w_\theta\}$, the average information gain on the state $\rho$ satisfies
	\begin{align}
		\sum_{\theta}w_\theta G(\mc{M}_\theta)_\rho=\braket{\rho|\hat{g}|\rho}\leq \big\lVert \hat{g}\big\rVert \cdot I_{\rm com}(\rho),  
		\label{gainsum}
	\end{align}
	where $\hat{g}=\sum_\theta w_\theta \hat{G}(\mc{M}_\theta)$ is the average view operator and $\lVert\cdot\rVert$ denotes the operator norm, i.e., the largest eigenvalue of an operator.
	
	\emph{Proof.} According to Eqs. (\ref{isom}, \ref{viewoperator}), for any density operator $\rho$ on $\mc{H}_d$ there is $\braket{\psi_d|\rho\rangle\langle\rho|\psi_d}=1/d$, $\braket{\rho|\rho}=tr(\rho^2)$ and $\braket{\rho|\tilde{M}_{i|\theta}}=p_{i|\theta}-tr(M_{i|\theta})/d$. Hence we have $\sum_{i,\theta}w_\theta\left(p_{i|\theta}-tr(M_{i|\theta})/d\right)^2=\braket{\rho|\hat{g}|\rho}\leq \lVert \hat{g}\rVert\cdot \braket{\rho|\mathbbm{1}_{\perp\psi_d}|\rho}=\lVert \hat{g}\rVert\cdot(tr(\rho^2)-1/d)$.
	
	Theorem 1 limits an observer's weighted average information gain over multiple measurements to be less than a proportion $\rVert\hat{g}\rVert$ of the complete information content \eqref{cominf} contained in quantum states. We show in Appendix. \ref{appview} that $\frac{1}{\Theta}\leq\rVert\hat{g}\rVert\leq1$ for a number  $\Theta$ of rank-1 projective measurements. To be more precise, for nondegenerate observables with one or more common eigenstates we have $\rVert\hat{g}\rVert=1$ and the rightmost side of  \myref{gainsum} is achieved by density operators whose eigenvectors corresponding to positive eigenvalues form a subset of the common eigenstates of observables, which means that no state-independent information exclusion exists. On the other hand, we have $\rVert\hat{g}\rVert=\max_\theta\{w_\theta\}\leq1$  for MUBs. Particularly,  for random measurements in one of  $\Theta$ MUBs, $w_1=\cdots=w_\Theta=\frac{1}{\Theta}$, thereby $\rVert\hat{g}\rVert=\frac{1}{\Theta}$. We therefore see that the average information gain is rather limited with an increasing number of MUBs.

 \emph{Example.} For random measurements on a qubit in one of two bases $\{\ket{i_1}\}$ and  $\{\ket{j_2}\}$, \myref{gainsum} gives $\braket{\rho|\hat{g}|\rho}\leq c_{\rm max}\cdot I_{\rm com}(\rho)$. Here, $c_{\rm max}=\max_ {i,j}\{\lvert\braket{i_1|j_2}\rvert^2\}$ is the maximal overlap between bases and in this simple example $1/2\leq c_{\rm max}\leq1$. By definition, $c_{\rm max}=1/2$ holds for MUBs, while for compatible bases  $c_{\rm max}=1$.

We remark that for those measurement strategies with which the associated view operator $\hat{g}\propto\mathbbm{1}_{\perp\psi_d}$, the rightmost side of   \myref{gainsum} can be achieved by any density operator on $\mc{H}_d$. Typical examples include random measurements in CMUBs, random selection of measurements from a complete set of mutually unbiased measurements \cite{kalev2014mum} and other design-structured POVMs \cite{ketterer2020design,rastegin2020designpovm,rastegin2023frame,renes2004sicpovm,gour2014,yoshida2022sicpovm} (see Appendix. \ref{appview} for details).

   \subsection{Information exclusion relations with memory}

\begin{figure}[b]
	\includegraphics[width=7.3cm]{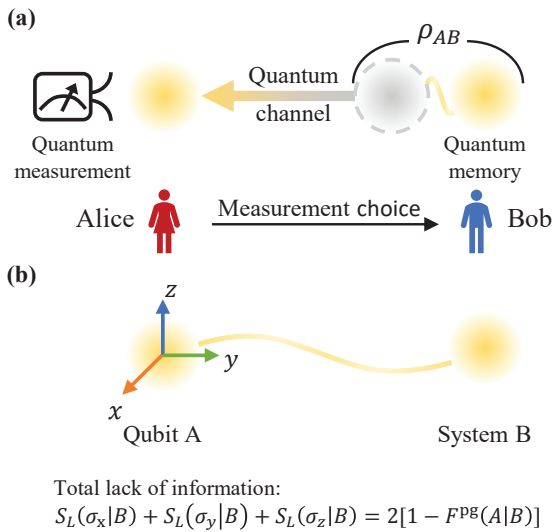}
	\caption{(a) Sketch of the proposal. (b) When Alice chooses to measure a qubit in one of three orthogonal directions, Bob's total lack of information (uncertainty) about Alice's measurement outcomes is negative linearly related to  the recoverable entanglement fidelity $F^{\rm pg}(A|B)$ of the initial state, which is time-invariant if there exits no information exchange with environments or between subsystems A and B.}
	\label{guess}
\end{figure}

	We move on to investigate the basic limits on an observer's information with respect to measurements on a distant quantum system, given access to another system (called memory). To illustrate, let us consider the guessing game \cite{berta2010}  involving two participants, Alice and Bob. As depicted in \myfig{guess}a, in the beginning, Bob prepares a bipartite system in the state $\rho_{AB}$, and sends subsystem A to Alice. Upon receiving subsystem A, Alice chooses a measurement according to the value $\theta$ of a random variable drawn from the probability distribution $\{w_\theta\}$, and announces her choice to Bob. Bob's win condition is to guess  the final state on Alice's side correctly.

   	To quantify Bob's lack of information about system A while possessing a memory system B, we define the conditional linear entropy as below
   \begin{align}
   	S_L(A|B)=1-d\cdot F^{\rm pg}(A|B).
   	\label{condentropy}
   \end{align}
	 Here, $F^{\rm pg}(A|B)=\frac{1}{d}tr\big\{\big[\big(\mathbbm{1}_A\otimes\rho_B^{-1/4}\big)\rho_{AB}\big(\mathbbm{1}_A\otimes\rho_B^{-1/4}\big)\big]^2\big\}$ is the recoverable entanglement fidelity with which $\rho_{AB}$ can be transformed into a maximally entangled state through the pretty good recovery operation on system B \cite{berta2014a,barnum2002}, and $d$ denotes the dimension of system A. In the case of a product state $\rho_{AB}=\rho_A\otimes\rho_B$, system B offers no side information about system A and \myref{condentropy} reduces to the linearized entropy $S_L(\rho_A)=1-tr(\rho_A^2)$, i.e., the complement of the information content \eqref{cominf} contained in the state $\rho_A$. More generally, according to the data-processing inequality \cite{muller2013,tomamichel2014} we have $S_L(A|B)\leq S_L(\rho_A)$, thereby a memory  helps to reduce Bob's ignorance. Further, $\rho_{AB}$ is necessarily entangled if  $S_L(A|B)<S_L(\sqrt{\rho_B})\equiv0$, since one's ignorance about the overall system in a separable state does not increase with the removal of any its local subsystem \cite{horodecki1996inf,horodecki1996}.

    For brevity, we will focus on rank-1 projective measurements. Bob has no direct access to system A once it is sent to Alice,  his understanding of the overall system when Alice chooses the $\theta$th measurement is described by the classical-quantum state
   \begin{align}
   	\rho_{\mc{M}_\theta B}=\sum_i|i\rangle\langle i|\otimes tr_A\left[ (M_{i|\theta}\otimes \mathbbm{1}_B)\rho_{AB}\right], \label{poststate}
   \end{align}
   where $M_{i|\theta}$ denotes the $i$th effect of the $\theta$th POVM $\mc{M}_\theta$ and $\{|i\rangle\langle i|\}$ are the measurement outcomes stored in a classical register. Then, the conditional linearized entropy \eqref{condentropy} evaluated on the classical-quantum state \eqref{poststate}, denoted $S_L(\mc{M}_\theta|B)=1-d\cdot F^{\rm pg}(\mc{M}_\theta|B)$, measures Bob's ignorance about Alice's measurement outcomes. Indeed, $ F^{\rm pg}(\mc{M}_\theta|B)$ is now precisely the probability for Bob to correctly guess Alice's measurement outcome by performing the pretty good measurement on system B \cite{hausladen1994,buhrman2008}.

 	\emph{Theorem 2.}  Suppose $\rho_{AB}$ describes a bipartite system and  $\{\mc{M}_{\theta}\}$ are rank-1 projective measurements on system A with selection probabilities $\{w_\theta\}$. The  average conditional linearized entropy is bounded below by
	\begin{align}
			\sum_{\theta}w_\theta & S_L(\mc{M}_\theta|B)\geq (1-\lVert \hat{g}\rVert)\cdot\big[1-F^{\rm pg}(A|B)\big].
		  \label{condunc}
		\end{align}

	We prove in Appendix. \ref{appthm2} a result that is valid for more general measurements. Like the memoryless IER \eqref{gainsum}, \myref{condunc} becomes an equality saturated by arbitrary bipartite state  if the equality  $\hat{g}=\lVert\hat{g}\rVert\cdot\mathbbm{1}_{\perp\psi_d}$ holds. Consequently, in the absence of information exchange with environments or between systems A and B, Bob's total information with respect to measurements on system A in CMUBs, as well as other design-structured measurements \cite{ketterer2020design,rastegin2020designpovm,rastegin2023frame,renes2004sicpovm,gour2014,yoshida2022sicpovm}, is time-invariant.

	Impressively, the r.h.s. of \myref{condunc} is a product of  two independent terms controlled by Alice and Bob respectively.  The first term,  $1-\lVert \hat{g}\rVert=:\mc{X}$, is a state-independent signature of information exclusion and Alice is free to manipulate it through her measurement strategy.  It varies in the range $\mc{X}\in[0,1-\frac{1}{\Theta}]$ when the number of observables under consideration is $\Theta$. To keep her measurement outcomes secret, Alice should avoid measuring observables that share a common eigenstate ($\mc{X}=0$), as Bob can completely eliminate his uncertainty by preparing system A precisely in that eigenstate. In contrast, Bob's uncertainty will be maximized if Alice randomly selects one of $\Theta$ MUBs ($\mc{X}=1-\frac{1}{\Theta}$). The special case when Alice chooses to measure the Pauli observables of a qubit is illustrated in  \myfig{guess}b. We need to mention here that a set of $\Theta$ MUBs may not exist for sufficiently large $\Theta$, and numerical methods can be utilized to maximize the exclusivity in such cases.

	On the other hand, the second term  decreases  monotonically with the recoverable entanglement fidelity $F^{\rm pg}(A|B)$ of the initial state $\rho_{AB}$.  Bob's pretty good guessing probability \cite{hausladen1994,buhrman2008} $ F^{\rm pg}(\mc{M}_\theta|B)$ would be less than 1 whenever $F^{\rm pg}(A|B)<1$. However, he can prepare an appropriate entangled state such that this fidelity enables him to guess the outcomes of measurements on system A with high probability.  Indeed, it is well known that maximally entangled states provide perfect side information. For example, two systems in the state $\ket{\psi_d}=\frac{1}{\sqrt{d}}\sum_{i=0}^{d-1}\ket{i}_A\otimes\ket{i}^*_B$  are perfectly correlated with no local information content at all, $I_{\rm com}(\rho_B)=I_{\rm com}(\rho_A)=0$, whereas the joint information content $I_{\rm com}(\rho_{AB})=1-1/d^2$ is maximal. This leads to  $F^{\rm pg}(A|B)=1$, namely, the correlation between A and B is strong enough to completely remove Bob's uncertainty.
	Just as is mentioned in \mycites{brukner1999,brukner2001}, the information content of a maximally entangled state is “exhausted in defining the joint properties” and “none is left for individual systems”.

    \section{Origin of tight WPDR${\bf s}$} \label{originWPDRs}
    
   	\begin{figure}[b]
    	\includegraphics[width=8.6cm]{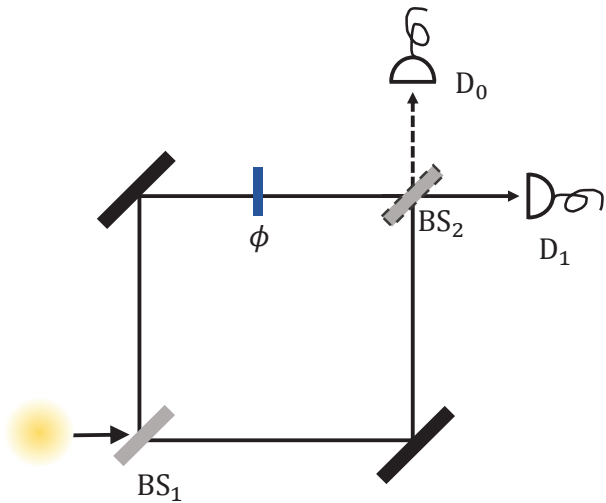}
    	\caption{Mach-Zehnder interferometer.  An input photon is directed into two paths by an asymmetric beam splitter (BS$_1$) and then is recombined on a 50:50 beam splitter (BS$_2$) to trigger two detectors (D).  Modulating the phase shift $\phi\in [0,2\pi]$ in the upper path, the phenomenon that the probability of click in each detector oscillates periodically reflects the interference pattern of path amplitudes, which is a signature of wave property. When BS$_2$ is removed, the photon behaves particle-like such that the local phase shift in path no longer affects the detection probability.  }
    	\label{mzi}
    \end{figure}
    
   We argue that the tight WPDRs are particular examples of the IERs \eqref{gainsum} and \eqref{condunc} for complementary observables. To see this, let us consider two complementary setups of the Mach-Zehnder interferometer depicted in \myfig{mzi}:  (i) the second beam splitter is removed to gain the path information of single photons inside the interferometer (let $\sigma^{\rm p}$ denote the associated path observable with binary outcomes “$+1$” and “$-1$”, corresponding to clicks in detectors $D_0$ and $D_1$ respectively );  (ii) BS$_2$ is inserted in and the phase shift $\phi$ is adjustable to reveal wave properties of photons (let $\sigma_\phi^{\rm w}$ denote the associated wave observable with binary outcomes “$\pm1$”).  It takes some calculation (see Appendix. \ref{appinter}) to see that \myref{gainsum} leads to the equality
	\begin{align}
		G(\sigma_\phi^{\rm w})_\rho&+G(\sigma_{\phi'}^{\rm w})_\rho=\cos(\phi'-\phi)\langle\sigma_\phi^{\rm w}\rangle\langle\sigma_{\phi'}^{\rm w}\rangle+\nonumber\\
		&[I_{\rm com}(\rho)-G(\sigma^{\rm p})_\rho]\sin^2(\phi'-\phi),
	\label{mziinfsum}
	\end{align}
where $\braket{\sigma}=tr(\sigma\rho)$ denotes the average of observable $\sigma$, and  $G(\sigma^{\rm p})_\rho=\frac{1}{2}\braket{\sigma^{\rm p}}^2$ and  $G(\sigma_\phi^{\rm w})_\rho=\frac{1}{2}\braket{\sigma_\phi^{\rm w}}^2$ are the respective information gains \eqref{infgain} for measuring the path and wave observables in the qubit $\rho$.

 \begin{figure}[b]
	\includegraphics[width=8.6cm]{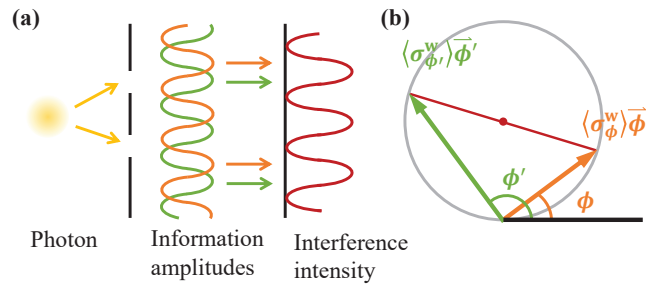}
	\caption{Interference pattern of information amplitude. (a) In the double-slit experiment, filtering out the component containing path information from photon's density operator, the remaining components lead to a fringe with 100\% contrast on the screen. The intensity varies periodically at different locations, corresponding to the intensity oscillation in the MZI as the phase shift  $\phi$ varies. (b) The information gains on single photons in the MZI when two complementary wave observables ($\phi'-\phi=\pi/2$) are measured constitute the complete description of the wave-like behavior. The fringe visibility is given by the diameter of the gray circle.}
	\label{interfere}
\end{figure}

	Observe that the information gain regarding an individual wave observable  oscillates as the phase shift $\phi$ varies,  \myref{mziinfsum} essentially depicts an interference pattern of the wave information. To make it clearer, let $\vec{\phi}$ and $\vec{\phi'}$ be two real unit vectors at an angle of $\phi'-\phi$.  \myref{mziinfsum} can then be restated as
	\begin{align}
		&\big|\langle\sigma_\phi^{\rm w}\rangle\vec{\phi}+e^{i\pi}\langle\sigma_{\phi'}^{\rm w}\rangle\vec{\phi'}\big|^2\nonumber\\
		=&2[I_{\rm com}(\rho)-G(\sigma^{\rm p})_\rho]\sin^2(\phi'-\phi).
		\label{wavepart}
	\end{align}
	It is interesting to note that the average values of wave observables behave like the “amplitudes of wave information” and interfere with each other, see \myfig{interfere}a. 
 Notably, the average interference intensity $\mc{I}=I_{\rm com}(\rho)-G(\sigma^{\rm p})_\rho$ on the r.h.s. of \myref{wavepart} disappears if the photon exhibits particle property only---the complete information content of $\rho$ is accessible through measuring the path observable or, formally, $I_{\rm com}(\rho)=G(\sigma^{\rm p})_\rho$. In this view, $\mc{I}=G(\sigma_\phi^{\rm w})_\rho+G(\sigma_{\phi+\pi/2}^{\rm w})_\rho$ (see the case $\phi'-\phi=\pm\frac{\pi}{2}$ in the preceding equation) emerges as a measure of wave property which can be determined by measuring two complementary wave observables.

 Conventionally, the wave property is frequently quantified by the fringe visibility \cite{greenberger1988,jaeger1993,jaeger1995,englert1996} 
	\begin{align}
		\mc{V}=\max\limits_{\phi}\big|p^0_\phi-p^1_\phi\big|,
		\label{fringe}
	\end{align}
  where $p_\phi^i$ is the probability that the $i$th detector clicks when the observable $\sigma^{\rm w}_\phi$ is measured.
We remark here that the average interference intensity is precisely half of the  fringe visibility squared, i.e., $\mc{V}=\max_\phi|\braket{\sigma^{\rm w}_\phi}|=\sqrt{\mc{I}/2}$ (see also \myfig{interfere}b for an illustration). Combined with the squared path distinguishability $\mc{D}^2=\braket{\sigma^{\rm p}}^2=2G(\sigma^{\rm p})_\rho$, we then arrive at the WPDR $\mc{V}^2+\mc{D}^2=2tr(\rho^2)-1$ \cite{qian2020}. We therefore see that the WPDR originates from the IER $\big[G(\sigma_\phi^{\rm w})_\rho+G(\sigma_{\phi+\pi/2}^{\rm w})_\rho\big]+G(\sigma^{\rm p})_\rho=I_{\rm com}(\rho)$ for three complementary observables, including the path observable and two wave observables with phase difference satisfying $\phi'-\phi=\pm\pi/2$.

\begin{figure}[b]
	\includegraphics[width=7.3cm]{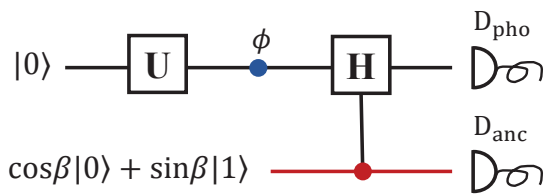}
	\caption{Delayed choice experiment controlled by an ancilla qubit (red line) in the state $\cos \beta\ket{0}+\sin \beta\ket{1}$. The second beam splitter in the two-way interferometer is now equivalently  represented by the Hadamard gate H: $\rm H \ket{0}\rightarrow(\ket{0}+\ket{1})/\sqrt{2}$ and $\rm H\ket{1}\rightarrow(\ket{0}-\ket{1})/\sqrt{2}$, while the first beam splitter, which can be asymmetric, is represented by a real unitary U: $\rm U\ket{0}\rightarrow\cos\alpha\ket{0}+\sin\alpha\ket{1}$ and $\rm U\ket{1}\rightarrow\sin\alpha\ket{0}-\cos\alpha\ket{1}$. }
	\label{delaychoice}
\end{figure}

	Theorem 1 applies also to the quantum delayed-choice experiment \cite{ionicioiu2011}, where complementary properties of photons are measured in a single experimental setup. As shown in \myfig{delaychoice}, the presence of BS$_2$  is controlled by an ancilla qubit, the value of which determines whether to reveal the wave property or particle property. In this case,  \myref{gainsum} limits an observer's weighted average information gain about  three complementary observables, with (unnormalized) weights  $w_1=\cos^2\beta$ for the path observable and  $w_2=w_3=\sin^2\beta$ for the wave observables. Therefore, the true nature of complementarity does not prohibit the observation of complementary properties in a single measurement setup, but necessarily restricts one's simultaneous information gain about them.

Another interesting issue concerns the WPDRs when an observer has side information about single photons in the Mach-Zehnder interferometer, but without direct access to them. Let us consider two photons in the bipartite state $\rho_{AB}$.
As a measure of information about photon A conditioned on photon B, we turn to the complement of the conditional linearized entropy \eqref{condentropy} below
	\begin{align}
		I(A|B)=d\cdot F^{\rm pg}(A|B)-1/d.
		\label{condinf}
	\end{align}
This is non-negative and reduces to the complete information of the reduced state $\rho_A$,  $I(A|B)=I_{\rm com}(\rho_A)$ when $\rho_{AB}=\rho_A\otimes\rho_B$ is a product state.

We derive in Appendix. \ref{appinter} the following generalization of \myref{wavepart}, 
	\begin{align}
		&tr[(\vec{\rho}_{\phi B}+e^{i\pi}\vec{\rho}_{\phi' B})^2]\nonumber\\
		=&2[I(A|B)-I(\sigma^{\rm p}|B)]\sin^2(\phi-\phi').
		\label{condwavepart}
	\end{align}
Here, $\vec{\rho}_{\phi B}=tr_A\big[(\rho_B^{-1/4}\rho_{AB}\rho_B^{-1/4})(\sigma_\phi^{\rm w}\otimes\mathbbm{1}_B)\big]\ \vec{\phi}$  is the “amplitude of conditional information” which connects to the conditional information $I(\sigma_\phi^{\rm w}|B)$ through its squared modulus  $tr(\vec{\rho}_{\phi B}^{\ 2})=2I(\sigma_\phi^{\rm w}|B)$.

	Equation \eqref{condwavepart} manifests the interference pattern of conditional information amplitude, with the r.h.s. of it being the interference intensity. Combining the average  intensity (wave property) $I(A|B)-I(\sigma^{\rm p}|B)=I(\sigma_\phi^{\rm w}|B)+I(\sigma^{\rm w}_{\phi+\pi/2}|B)$ with the conditional which-way information (particle property) $I(\sigma^{\rm p}|B)$, we then obtain the WPDR
	$\big[I(\sigma_\phi^{\rm w}|B)+I(\sigma_{\phi+\pi/2}^{\rm w}|B)\big]+I(\sigma^{\rm p}|B)=I(A|B)$.  Again,  we see that a tight WPDR saturated by all bipartite systems with dimension $d_A=2$ arises from an IER for three complementary observables, wherein two complementary wave observables constitute the complete description of wave property.

     \section{Discussions}\label{discussions}

	Our theory for characterizing information complementarity from a measurement-based perspective enables us to analyze the behaviors of quantum systems through their manifestations in versatile measurement setups, without delving into the exhaustive calculations with quantum state parameters. As two examples, we explore the implications of our IERs (\ref{gainsum}, \ref{condunc}) for entanglement detection and EURs respectively.  
	
\subsection{Entanglement detection}

   Quantum correlation tends to suppress the local information content contained in individual subsystems.  For example, a pair of maximally entangled qubits possess only joint properties in the sense that each single qubit is in the completely mixed state. We introduce the correlation measure $J(\rho_{AB})=\sum_{i,\theta} w_\theta|tr(J_{i|\theta}\cdot\rho_{AB})|$ for local measurements $\{\mc{M}^A_{\theta}\otimes \mc{M}^B_{\theta}\}$ on individual copies of the bipartite state $\rho_{\rm AB}$, where $M_{i|\theta}$ denotes the $i$th effect of the $\theta$th measurement and
	$J_{i|\theta}=(M^A_{i|\theta}-\frac{1}{d_A}tr(M^A_{i|\theta})\mathbbm{1}_{A})\otimes (M^B_{i|\theta}-\frac{1}{d_B}tr(M^B_{i|\theta})\mathbbm{1}_{B})$ are the correlation detection operators. We show in Appendix. \ref{appthm3} the following.

	\emph{Theorem 3.} For any bipartite separate state $\rho_{AB}$, it holds that
	\begin{align}
		J(\rho_{AB})\leq \sqrt{L_A\cdot L_B},
		\label{entangleineq}
	\end{align}
	where $L=\lVert\hat{g}\rVert(1-1/d)$  is the state-independent upper bound on local information gain given by \myref{gainsum}.

	\begin{figure}[t]	\includegraphics[width=8.6cm]{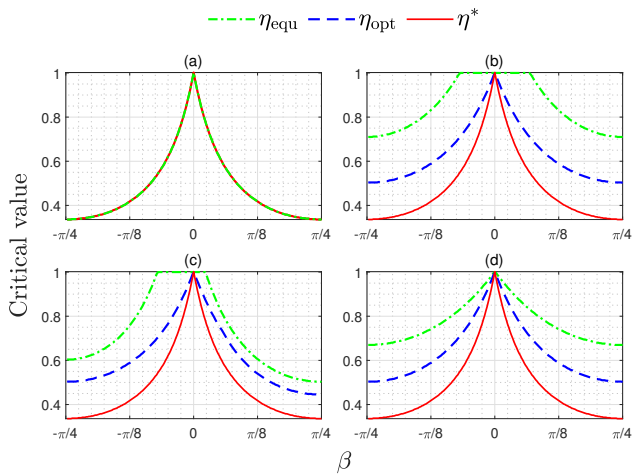}
		\caption{Numerical comparison of the critical value of $\eta$ for the state \eqref{estate} to be entangled (denoted $\eta^*$) and that to violate \myref{entangleineq} under four different choices of three local observables, with equal weights and optimized weights respectively (denoted $\eta_{\rm equ}$ and $\eta_{\rm opt}$). The three local observables considered here are $\sigma_y\otimes\sigma_y, \sigma_z\otimes\sigma_z$ and (a) $\sigma_x\otimes\sigma_x$. (b) $\sigma_z\otimes\sigma_x$. (c) $(\frac{1}{2}\sigma_z+\frac{\sqrt{3}}{2}\sigma_x)\otimes (\frac{1}{2}\sigma_z+\frac{\sqrt{3}}{2}\sigma_x)$. (d) $\sigma_z\otimes\sigma_z$.
			\label{entangle}}
	\end{figure}

	Consequently, a violation of \myref{entangleineq} necessarily indicates the presence of entanglement.   
	As a concrete example,  we can apply \myref{entangleineq} to the mixture of a pure two-qubit state $\ket{\psi(\beta)}=\cos\beta\ket{00}+\sin\beta\ket{11}\ (-\frac{\pi}{4}\leq\beta\leq\frac{\pi}{4})$ and white noise:
	\begin{align}
		\rho_{\eta,\beta}=\eta|\psi(\beta)\rangle\langle\psi(\beta)|+(1-\eta)\mathbbm{1}_4/4,\ (0\leq\eta\leq1).
		\label{estate}
	\end{align}
	Note that the noiseless state $\ket{\psi(\beta)}$ is entangled as long as $\beta\neq0$.
	Now the question is how much noise it can resist from being separable, i.e., the critical value $\eta^*$ of $\eta$ below which $\rho_{\eta,\beta}$ ceases to be entangled.
	In \myfig{entangle}, we present numerical results regarding the critical values $\eta_{\rm equ}$ and $\eta_{\rm opt}$ for the state \eqref{estate} to violate  \myref{entangleineq}, under measurements with equal weights and optimized weights respectively.
	As depicted, three complementary observables with equal weights are enough to detect all the entanglement $(\eta_{\rm equ}=\eta_{\rm opt}=\eta^*)$. For more general observables $\eta_{\rm opt}\leq\eta_{\rm equ}$, an optimization over the weights $\{w_\theta\}$ yields better performance.

	\subsection{Implications for EURs}
	
	Entropic uncertainty relations (EURs) that take  into account information leakage from a memory system play a crucial role in various aspects of quantum information processing \cite{coles2017}, particularly in the security analysis of quantum protocols \cite{berta2010}.  However, existing EURs \cite{coles2017} are thus far limited since they are restricted to providing lower bounds on simply entropy sums.   On a conceptual level, there is no reason to assign equal weights, instead of biased weights, to different measurements. Based on Theorem 2, we have the following lower bounds on the weighted sum of entropies over multiple measurements (see the proof in Appendix, \ref{appthm4}).

	\emph{Theorem 4.}   Suppose $\rho_{AB}$ describes a bipartite system and  $\{\mc{M}_{\theta}\}$ are rank-1 projective measurements to be performed on system A with selection probabilities $\{w_\theta\}$. The smooth minimum entropy evaluated on the state \eqref{poststate} satisfies $\sum_\theta w_\theta H_{\rm min}^{\varepsilon}(\mc{M}_\theta|B)\geq q_{\rm min}^\varepsilon$, where
	\begin{align}
		q_{\rm min}^\varepsilon=-\log \big[\lVert\hat{g}\rVert+F^{\rm pg}(A|B)(1-\lVert\hat{g}\rVert)\big]-\log\frac{2}{\varepsilon^2}.
		\label{collentropy}
	\end{align}

    The conditional smooth minimum entropy \cite{renner2008security} (see also \mycite{coles2017}) is a fundamental tool for the security analysis of quantum protocols. In quantum cryptographic protocols where an eavesdropper aims to know an experimenter's measurement outcomes by probing a memory system, the weighted EURs we introduced provide guidance for adjusting the probabilities of selecting distinct measurements to minimize potential information leakage. It is conceivable that equal selection probabilities are not optimal for biased measurements. Optimized selection probabilities are thus crucial for elaborating the measurement strategies to enhance security and achieve stronger levels of protection. Importantly, this optimization does not require additional quantum costs and can be easily done on a classical computer.

   \section{Conclusion}\label{conclusion}

    In summary, we have developed a general approach to formulate the complementarity principle quantitatively in terms of basic limits on one's ability to gain information on quantum systems under versatile measurement setups, with and without memory respectively. 
   Our framework sheds new light on the interpretation of wave-particle duality for single photons in the two-way interferometry experiments from an information-theoretical perspective. Extending this interpretation to multi-path interferometers presents an intriguing avenue for future investigation. Moreover, our IERs have direct applications in certifying genuine quantum features of physical systems, such as entanglement detection based on local measurement outcomes. An extended form of EURs can also be derived from our IERs, which could offer practical advantages in various quantum information processing.

   \acknowledgments
 
 This work is supported by the National Natural Science Foundation of China (Grants No. 12175104 and No. 12274223), the Innovation Program for Quantum Science and Technology (2021ZD0301701), the Natural Science Foundation of Jiangsu Province (No. BK20211145), the Fundamental Research Funds for the Central Universities (No. 020414380182), the Key Research and Development Program of Nanjing Jiangbei New Area (No. ZDYD20210101),  the Program for Innovative Talents and Entrepreneurs in Jiangsu (No. JSSCRC2021484), and the Program of Song Shan Laboratory (Included in the management of Major  Science and Technology Program of Henan Province) (No. 221100210800-02).
	
    \appendix
	
    \section{View operator and properties} \label{appview}
    
	Consider a set of  POVMs $\{M_{i|\theta}\}$ assigned with weights $\{w_\theta\}$ $(w_\theta\geq0,\ \sum_\theta w_\theta=1)$. We define the associated average view operator to be
	\begin{align}
		\hat{g}=\sum_{i,\theta}w_\theta\hat{G}(\mc{M}_\theta)=\sum_{i,\theta}w_\theta|\tilde{M}_{i|\theta}\rangle\langle \tilde{M}_{i|\theta}|,
		\label{averageview}
	\end{align}
	where $\tilde{M}_{i|\theta}=M_{i|\theta}-\frac{1}{d}tr(M_{i|\theta})\mathbbm{1}_d$ is traceless, or equivalently, $\ket{\tilde{M}_{i|\theta}}=\ket{M_{i|\theta}}-|\psi_d\rangle\langle\psi_d|M_{i|\theta}\rangle$ is orthogonal to $\ket{\psi_d}$. View operators are positive semi-definite,  $\hat{G}\geq0$ on the $(d^2-1)$-dimensional subspace $\mc{H}_{\perp\psi_d}$ of $\mc{H}_d^{\otimes2}$ orthogonal to $\ket{\psi_d}$, and vanish for trivial POVMs whose effects satisfy $M_{i|\theta}=\frac{1}{d}tr(M_{i|\theta})\mathbbm{1}_d$.

The matrix representation of  $\hat{g}$ under an orthonormal basis $\{\ket{a}\}$ of $\mc{H}_{\perp\psi_d}$ takes the form
	\begin{align}
		g_{a,a'}=&\sum_{i,\theta}w_\theta\langle a|\tilde{M}_{i|\theta}\rangle\langle \tilde{M}_{i|{\theta}}|a'\rangle=	(RR^+)_{a,a'}.
 		\label{gram}
	\end{align}
Here, the matrix elements of $R$ are given by $R_{a,b(i,\theta)}=\sqrt{w_\theta}\langle a|\tilde{M}_{i|\theta}\rangle$, with $b$  being a bijection from the labels $\{(i,\theta)\}$ of POVM effects to the labels $\{a\}$ of the basis vectors $\{\ket{a}\}$. Note that the positive  eigenvalues of $g=RR^+$ are identical to those of  the Gram matrix for the vectors $\{\sqrt{w_\theta}\ket{\tilde{M}_{i|\theta}}\}$, that is, $\bar{g}=R^+R$. To obtain eigenvalues of a view operator $\hat{g}$, it will be enough to deal with the Gram matrix  $\bar{g}$, whose elements are $\bar{g}_{b(i,\theta),b(j,\theta')}=\sqrt{w_\theta w_{\theta'}}\langle \tilde{M}_{i|\theta}|\tilde{M}_{j|{\theta'}}\rangle$. 

\emph{Claim 1.} POVMs that form a design structure are mutually complementary.

\emph{Claim 2.}  The combined view operator associated with a complete set of design-structured POVMs is proportional to the identity operator on $H_{\perp\psi_d}$.

\emph{Claim 3.} The average view operator of a set of MUBs with weights $\{w_\theta\}$ satisfies  $\lVert\hat{g}\rVert=\max_\theta\{w_\theta\}$.

\emph{Proof.} Design-structured measurements include complete sets of mutually unbiased measurements (MUMs) \cite{kalev2014mum}, general symmetric-informationally-complete POVMs \cite{renes2004sicpovm,gour2014,yoshida2022sicpovm}, POVMs from equiangular tight frames  \cite{rastegin2023frame} and POVMs from general quantum designs \cite{ketterer2020design,rastegin2020designpovm}.
Without loss of generality, we prove the above claims for MUMs. MUMs \cite{kalev2014mum} are  $d$-outcome POVMs satisfying $tr(M_{i|\theta})=1$, $tr(M_{i|\theta}M_{j|\theta'})=\frac{1}{d}$, and $tr(M_{i|\theta}M_{j|\theta})=\kappa\delta_{ij}+\frac{1-\kappa}{d-1}(1-\delta_{ij})$ for all $i,j=0,\cdots,d-1$ and $\theta\neq\theta'$.  Here  $\kappa\in(\frac{1}{d},1]$ is called the efficiency parameter, wherein $\kappa=1$ corresponds to projective measurements in MUBs \cite{ivonovic1981,wootters1989,klappenecker2004,pittenger2004}.

Consider the view operator $\hat{G}_{\rm mum}=\sum_{\theta}\hat{G}(\mc{M}_\theta)$ associated with a set of MUMs \cite{kalev2014mum} on $\mc{H}_d$, according to \myref{gram} the corresponding Gram matrix $\bar{G}$ is given as
\begin{align}
	&\bar{G}_{b(i,\theta),b(j,\theta')}=\braket{\tilde{M}_{i|\theta}|\tilde{M}_{j|{\theta'}}}=tr(M_{i|\theta}M_{j|{\theta'}})-\frac{1}{d} \nonumber\\
 =&\delta_{\theta\theta'}\Big[\frac{\kappa d-1}{d-1}\delta_{ij}+\frac{1-\kappa d}{d(d-1)}\Big].\label{grammum}
\end{align}
According to \myref{grammum}, obviously two MUMs are complementary since $\hat{G}(\mc{M}_\theta)\cdot\hat{G}(\mc{M}_{\theta'})=0$ whenever $\theta\neq\theta'$. Next, let us focus on the $d\times d$ submatrix 
\begin{align}
	\bar{G}_{b(i,1),b(j,1)}=\frac{\kappa d-1}{d-1}\mathbbm{1}_d-\frac{\kappa d-1}{d(d-1)}Q, \label{submatrix}
\end{align}
where $Q$ denotes the matrix satisfying $Q_{i,j}=1$ for all $i,j=0,\cdots,d-1$. This submatrix \eqref{submatrix} has $d-1$ identical nonzero eigenvalues $(\kappa d-1)/(d-1)$, thus the view operator of a complete set of  $d+1$ MUMs (CMUMs) has $(d+1)(d-1)=d^2-1$ identical nonzero eigenvalues. In other words,  $\hat{G}_{\rm cmum}=\frac{\kappa d-1}{d-1}\mathbbm{1}_{\perp\psi_d}$, with $\mathbbm{1}_{\perp\psi_d}=\mathbbm{1}_{d\times d}-|\psi_d\rangle\langle\psi_d|$ being the identity operator on the $(d^2-1)$-dimensional space $H_{\perp\psi_d}$.  Claim 3 follows  from the fact that MUBs (i.e., MUMs with efficient parameter $\kappa=1$) are complementary, thus  $\lVert\hat{g}\rVert=\max_\theta\{ w_\theta \lVert\hat{G}(\mc{M}_\theta)\rVert\}=\max_\theta\{ w_\theta\}$.

\emph{Claim 4.} For arbitrary $d$-outcome POVMs $\mc{M}=\{M_i\}$ on $\mc{H}_d$  that consists of  equal-trace effects (ETE-POVMs), i.e., $tr(M_{0})=\cdots=tr(M_{d-1})$, we have $\lVert\hat{G}(\mc{M})\rVert\leq1$.

\emph{Claim 5.} For any set of  $d$-outcome ETE-POVMs $\{M_\theta\}$ on $\mc{H}_d$,
$\lVert\hat{g}\rVert= \Big\lVert \sum_\theta w_\theta\hat{G}(\mc{M}_\theta)\Big\rVert\leq \sum_\theta w_\theta\left\lVert\hat{G}(\mc{M}_\theta)\right\rVert\leq1
$.

\emph{Claim 6.} For a number $\Theta$ of $d$-outcome ETE-POVMs $\{M_\theta\}$ on $\mc{H}_d$ with equal weights,  $\lVert\hat{g}\rVert=\frac{1}{\Theta}\lVert\sum_\theta\hat{G}(\mc{M}_\theta)\rVert=1$ iff the overlap matrix $W$, defined as $W_{b(i,\theta),b(j,\theta')}=tr(M_{i|\theta}M_{j|{\theta'}})$, is reducible.

\emph{Proof.} Consider the Gram matrix $\bar{G}_{i,j}=\langle \tilde{M}_{i}|\tilde{M}_{j}\rangle=tr(M_iM_j)-\frac{1}{d}$. We can rewrite it as $\bar{G}=W-Q/d$, where $W_{i,j}=tr(M_iM_j)$ is referred to as the overlap matrix, and $Q_{i,j}=1$ for all $i,j$. Note $W$ is doubly stochastic, i.e., $\sum_iW_{i,j}=\sum_jW_{i,j}=1$,  its first eigenvalue (arranged in descending order) must be $\lambda_1(W)=1$. Moreover, the corresponding eigenvector $v_1=(1,\cdots,1)^{\rm T}$ is also an eigenvector of  $Q$ which corresponds to the unique nonzero eigenvalue $d$ of $Q$. Immediately  $\bar{G}\cdot v_1=0$, and $\lVert\hat{G}(\mc{E})\rVert=\lVert\bar{G}\rVert=\lambda_2(W)\leq\lambda_1(W)=1$.
Claim 5 follows directly from Claim 4. Further, considering  that the matrix $\frac{1}{\Theta}W$ is doubly stochastic, according to Theorem 3.1 of \mycite{fiedler1972} we have $\lambda_2(\frac{1}{\Theta}W)=1$ iff $W$ is reducible.

\section{Proof of Theorem 2} \label{appthm2}

Let $\{M_{i|\theta}\}$ be a set of generalized measurements such that the POVM effects of each measurement are equal-trace, i.e., $tr(M_{0|\theta})=\cdots=tr(M_{l_\theta-1|\theta})=d/l_\theta$, where $l_\theta$ denotes the number of effects in the $\theta$th POVM.  After Alice performed the $\theta$th measurement on system A, Bob's understanding of the overall system is then described by the classical-quantum state 
\begin{align}
    \rho_{\mc{M}_\theta B}=\sum_{i=0}^{l_\theta-1}|i\rangle\langle i|\otimes (K_{i|\theta}\otimes\mathbbm{1}_B)\rho_{AB}(K_{i|\theta}^+\otimes\mathbbm{1}_B).
    \label{cqstate}
\end{align}
Here, $\{K_{i|\theta}\}$ are the Kraus operators \cite{kraus1983} which satisfy $K_{i|\theta}^+K_{i|\theta}=M_{i|\theta}$ by definition.

    To prove Theorem 2, we only need to show the operator 
\begin{align}
 &\hat{\Gamma}_{AB}=\lVert \hat{g}\rVert \mathbbm{1}_A\otimes\rho_B^{1/2}+\Big(\sum_{\theta}\frac{w_\theta }{l_\theta}-\frac{1}{d}\lVert \hat{g}\rVert\Big)\bar{\rho}_{AB}-\nonumber\\
 \sum_{\theta,i,x,x'}&w_\theta K_{i|\theta}^+|x\rangle_A\langle x'|K_{i|\theta}\otimes{}_A\langle x| K_{i|\theta}\bar{\rho}_{AB}K_{i|\theta}^+|x'\rangle_A
 \label{gamma}
\end{align}
 is positive semi-definite on the space $\mc{H}_A\otimes\mc{H}_B$, where $\{\ket{x}\}_{x=0}^{d-1}$ is an orthonormal basis of $\mc{H}_A$ and $\bar{\rho}_{AB}=\big(\mathbbm{1}_A\otimes\rho_B^{-1/4}\big)\rho_{AB}\big(\mathbbm{1}_A\otimes\rho_B^{-1/4}\big)$. Notice that the measurement-induced local transformation $\rho_{AB}\rightarrow\rho_{\mc{M}_\theta B}$ commutes with the map $\rho_{AB}\rightarrow \bar{\rho}_{AB}$, from  $\hat{\Gamma}_{AB}\geq0$ we have
\begin{align}
	&tr(\hat{\Gamma}_{AB}\bar{\rho}_{AB})=\lVert \hat{g}\rVert+\Big(\sum_{\theta}w_\theta/l_\theta-\lVert \hat{g}\rVert/d\Big)tr(\bar{\rho}_{AB}^2)\nonumber\\
  \geq&\sum_{i,\theta}w_\theta tr\left[K_{i|\theta}\bar{\rho}_{AB}K_{i|\theta}^+K_{i|\theta}\bar{\rho}_{AB}K_{i|\theta}^+\right]\nonumber\\
  =&\sum_{\theta}w_\theta \cdot tr(\bar{\rho}_{\mc{M}_\theta B}^2),
\label{gammaAB}
\end{align}
where $\bar{\rho}_{\mc{M}_\theta B}=(\mathbbm{1}_A\otimes\rho_B^{-1/4})\rho_{\mc{M}_\theta B}(\mathbbm{1}_A\otimes\rho_B^{-1/4})$. This leads us to
\begin{align}
		\sum_{\theta}w_\theta & S_L(\mc{M}_\theta|B)\geq 1-\lVert \hat{g}\rVert  \nonumber\\
  -&\big(\sum_{\theta}w_\theta/l_\theta-\lVert \hat{g}\rVert/d\big)\big[1-S_L(A|B)\big].
  \label{conduncsum}
	\end{align}
In the case of rank-1 projective measurements, $l_1=\cdots=l_\Theta$ are equal to the dimension $d$ of system A. With \myref{conduncsum}  Theorem 2 is already obvious.

Next, we proceed to show  $\hat{\Gamma}\geq0$. Observe the operator below is positive semi-definite
\begin{align}
	&\hat{\Omega}=\lVert \hat{g}\rVert\cdot(\mathbbm{1}_{d}^{\otimes 2}-|\psi_d\rangle\langle\psi_d|)-\hat{g}
=\lVert \hat{g}\rVert\cdot\mathbbm{1}_{d}^{\otimes2}
\label{omega}\\
 -&\sum_{i,\theta}w_\theta|M_{i|\theta}\rangle\langle M_{i|\theta}|+(\sum_\theta w_\theta d/l_\theta-\lVert \hat{g}\rVert)|\psi_d\rangle\langle\psi_d|\geq0,\nonumber
\end{align}
 and, accordingly,  so does its partial transpose over the second space
\begin{align}
  \hat{\Omega}^{T_2}=\lVert \hat{g}\rVert\cdot\mathbbm{1}_{d}^{\otimes 2}&-\sum_{i,\theta}w_\theta(|M_{i|\theta}\rangle\langle M_{i|\theta}|)^{T_2}\hspace{1cm}\nonumber\\
 +&(\sum_\theta w_\theta d/ l_\theta-\lVert \hat{g}\rVert)\hat{F}\geq0.
\end{align}
In the above
\begin{align}
    \hat{F}=(|\psi_d\rangle\langle\psi_d|)^{T_2}=\frac{1}{d}\sum_{i,j=0}^{d-1}|i\rangle\langle j|\otimes |j\rangle\langle i|,
\end{align}
and
\begin{align}
&(|M_{i|\theta}\rangle\langle M_{i|\theta}|)^{T_2}\nonumber\\
=&d(K^+_{i|\theta}K_{i|\theta}\otimes\mathbbm{1}_d|\psi_d\rangle\langle \psi_d|K^+_{i|\theta}K_{i|\theta}\otimes\mathbbm{1}_d)^{T_2}\\
=&\sum_{x,x'}K_{i|\theta}^+|x\rangle\langle x'|K_{i|\theta}\otimes\sum_{y,y'}(K_{i|\theta})_{xy}|y'\rangle\langle y| (K^+_{i|\theta})_{y'x'}.\nonumber
\end{align}

Let $\hat{\Omega}_{AC}$ be the operator $\hat{\Omega}$ when defined on the space $\mc{H}_A\otimes \mc{H}_C$. Similarly,  $\rho_{CB}$ and $\rho_{AB}$ denote the same density operator $\rho$ but defined on different spaces. Then, with $T_C$ denoting to the partial transpose over the space $\mc{H}_C$, it can be checked that
\begin{align}	
\hat{\Gamma}_{AB}=tr_C(\hat{\Omega}_{AC}^{T_C}\bar{\rho}_{CB}).
\end{align}
As a positive semi-definite Hermitian operator,  $\hat{\Omega}$ can be written as the sum of (unnormalized) rank-1 projectors
$\hat{\Omega}=\sum_x|\pi_x\rangle\langle\pi_x|$, thereby
\begin{align}
	&\hat{\Gamma}_{AB}=\sum_xtr_C\Big[(|\pi_x\rangle_{AC}\langle\pi_x|)^{T_C}\bar{\rho}_{CB}\Big] \nonumber\\
	=&\sum_xtr_C\Big[\sqrt{\bar{\rho}_{CB}}(|\pi_x\rangle_{AC})^{T_C}({}_{AC}\langle\pi_x|)^{T_C}\sqrt{\bar{\rho}_{CB}}\Big]\\
	= &\sum_xtr_C\big(\Pi_x^+\Pi_x\big), \nonumber
\end{align}
where $\Pi_x=\sqrt{\bar{\rho}_{CB}}(|\pi_x\rangle_{AC})^{T_C}$. Considering that  $\Pi_x^+\Pi_x\geq0$  are positive semi-definite operators on the space $\mc{H}_A\otimes\mc{H}_B\otimes\mc{H}_C$, immediately we have $\hat{\Gamma}_{AB}\geq0$. This completes the proof of Theorem 2.

For design-structured measurements, the corresponding combined view operators are proportional to $\mathbbm{1}_{\perp\psi_d}=\mathbbm{1}_{d}^{\otimes 2}-|\psi_d\rangle\langle\psi_d|$, then $\hat{\Omega}=\hat{\Gamma}_{AB}=0$ and \myref{gammaAB} becomes an equality saturated by arbitrary state $\rho_{AB}$ on $\mc{H}_A\otimes\mc{H}_C$.

\section{Interference pattern of information amplitude} \label{appinter}

We denote by $\{\ket{i_\phi}\}_{i=0,1}$  the measurement basis with respect to the experimental setup where BS$_2$ of the two-way interferometer (see \myfig{mzi}) is inserted in and the phase shift is $\phi$. Then, the  associated view operator is 
\begin{align}
& \hat{G}_\phi^{\rm w}=\sum_{i=0,1}|i_\phi\rangle\langle i_\phi|\otimes|i_\phi\rangle^*\langle i_\phi|-|\psi_2\rangle\langle\psi_2| \label{dualityviewu}\\
 =&\frac{1}{2}\sum_{i,j=0,1}(-1)^{i+j}|i_\phi\rangle\langle j_\phi|\otimes|i_\phi\rangle^*\langle j_\phi|=\frac{1}{2}|\sigma_\phi^{\rm w}\rangle\langle\sigma_\phi^{\rm w}|,\nonumber
 \end{align}
 where  $\ket{\sigma^{\rm w}_\phi}=|0_\phi\rangle\otimes |0_\phi\rangle^*-|1_\phi\rangle\otimes |1_\phi\rangle^*$ is the vector representation of the wave observable $\sigma^{\rm w}_\phi=|0_\phi\rangle\langle0_\phi|-|1_\phi\rangle\langle1_\phi|$  given by the isomorphism \eqref{isom}. Similarly, the view operator associated with the path observable $\sigma^{\rm p}$ is given as $\hat{G}^{\rm p}=\frac{1}{2}|\sigma^{\rm p}\rangle\langle\sigma^{\rm p}|$.

 Recall that the path observable is complementary to wave observables and, consequently, the view operators $\{\hat{G}^{\rm p},\hat{G}^{\rm w}_\phi,\hat{G}^{\rm w}_{\phi+\frac{\pi}{2}}\}$ are mutually orthogonal and satisfy
 \begin{align}
 	\hat{G}^{\rm p}+\hat{G}^{\rm w}_\phi+\hat{G}^{\rm w}_{\phi+\frac{\pi}{2}}\equiv\mathbbm{1}_{\perp\psi_2}.
 	\label{identity}
 \end{align}

Moreover, for arbitrary two wave observables $\sigma_{\phi'}^{\rm w}$ and $\sigma_\phi^{\rm w}$, it can be easily checked that
 \begin{align}
     \ket{\sigma_{\phi'}^{\rm w}}=\cos(\phi'-\phi)\ket{\sigma_\phi^{\rm w}}+\sin(\phi'-\phi)\ket{\sigma_{\phi+\frac{\pi}{2}}^{\rm w}}.
 \end{align}
 This leads us to
  \begin{align}
&\sin^2(\phi'-\phi)\hat{G}_{\phi+\frac{\pi}{2}}^{\rm w}=\sin(\phi'-\phi)^2|\sigma_{\phi+\frac{\pi}{2}}^{\rm w}\rangle\langle \sigma_{\phi+\frac{\pi}{2}}^{\rm w}|\nonumber\\
=&\hat{G}_{\phi'}^{\rm w}+\hat{G}_\phi^{\rm w}\hat{G}_{\phi'}^{\rm w}\hat{G}_\phi^{\rm  w}-\hat{G}_\phi^{\rm w}\hat{G}_{\phi'}^{\rm w}-\hat{G}_{\phi'}^{\rm w}\hat{G}_\phi^{\rm w}.
\label{orthogonalize}
 \end{align}
Combining equations \eqref{orthogonalize} and \eqref{identity} we have for any qubit density operator $\rho$
\begin{align}
    &\sin^2(\phi'-\phi)\braket{\rho|\hat{G}^{\rm p}+\hat{G}_\phi^{\rm w}+\hat{G}_{\phi+\frac{\pi}{2}}^{\rm w}|\rho} \nonumber\\
    =&\sin^2(\phi'-\phi)[G(\sigma^{\rm p})_\rho+G(\sigma^{\rm w}_\phi)_\rho]+G(\sigma_{\phi'}^{\rm w})_\rho\nonumber\\
    &+\cos^2(\phi'-\phi)G(\sigma_\phi^{\rm w})_\rho -\cos(\phi'-\phi)\braket{\sigma_\phi^{\rm w}}\braket{\sigma_{\phi'}^{\rm w}}\nonumber\\
   =&\sin^2(\phi'-\phi)\braket{\rho|\mathbbm{1}_{\perp\psi_2}|\rho}
   =\sin^2(\phi'-\phi)I_{\rm com}(\rho),
\end{align}
which completes the proof of \myref{mziinfsum}.

To derive the interference pattern of the “amplitude of conditional information” as given in  \myref{condwavepart}, let us consider the equality
\begin{align}
  \hat{\Delta}_{AC}=&\hat{G}^{\rm w}_\phi+\hat{G}_{\phi'}^{\rm w}-\hat{G}_\phi^{\rm w}\hat{G}_{\phi'}^{\rm w}
  -\hat{G}_{\phi'}^{\rm w}\hat{G}_\phi^{\rm w}   \nonumber\\
  =&\sin^2(\phi'-\phi)[\mathbbm{1}_{\perp\psi_2}-\hat{G}^{\rm p}].
\end{align}
From the proof of Theorem 2 we have
\begin{align}
&tr_{ABC}(\bar{\rho}_{AB}\hat{\Delta}_{AC}^{T_C}\bar{\rho}_{CB})=\sin^2(\phi'-\phi)[1-tr(\bar{\rho}_{\sigma^{\rm p}B}^2)] \nonumber\\
=&tr(\bar{\rho}_{\sigma^{\rm w}_\phi B}^2)+tr(\bar{\rho}_{\sigma^{\rm w}_{\phi'}B}^2)-tr(\bar{\rho}_{AB}^2)
-T(\phi'-\phi) ,
\label{apppw}
\end{align}
where $\rho_{\sigma B}$ denotes the classical-quantum state \eqref{cqstate} after measuring the observable $\sigma$ and
\begin{widetext}
\begin{align}
   &T(\phi'-\phi)=2\ tr_{ABC}\big[\bar{\rho}_{CB}\bar{\rho}_{AB}(\hat{G}_\phi^{\rm w}\hat{G}_{\phi'}^{\rm w})^{T_C}_{AC}\big]=2\cos(\phi'-\phi)tr_{ABC}\big[\bar{\rho}_{CB}\bar{\rho}_{AB}(|\sigma_\phi^{\rm w}\rangle\langle\sigma_{\phi'}^{\rm w}|)^{T_C}_{AC}\big]\nonumber\\
   =&\cos(\phi'-\phi)\sum_{i,j=0,1}(-1)^{i+ j}\cdot tr_{ABC}\ \big[\bar{\rho}_{AB}|i_\phi\rangle_A\langle j_{\phi'}|\otimes |j_{\phi'}\rangle_C\langle i_\phi|\bar{\rho}_{CB}\big]\nonumber\\
   =&\cos(\phi'-\phi)\sum_{i,j=0,1}(-1)^{i+ j}\cdot tr_B\big[{}_A\langle j_{\phi'}|\bar{\rho}_{AB}|i_\phi\rangle_A\langle i_\phi|\bar{\rho}_{AB}|j_{\phi'}\rangle_A\big]\nonumber\\
   =&\cos(\phi'-\phi)tr_{AB}(\sigma_{\phi'}^{\rm w}\bar{\rho}_{AB}\sigma_\phi^{\rm w}\bar{\rho}_{AB})\nonumber\\
    =&\cos(\phi'-\phi)tr_B[tr_A(\bar{\rho}_{AB}\sigma_\phi^{\rm w}\otimes\mathbbm{1}_B)tr_A(\bar{\rho}_{AB}\sigma_{\phi'}^{\rm w}\otimes\mathbbm{1}_B)]-\cos^2(\phi'-\phi)[tr(\bar{\rho}^2_{AB})-1].
\end{align}

Observe \myref{apppw} can be rewritten as
\begin{align}
    &tr(\bar{\rho}_{\sigma^{\rm w}_\phi B}^2)-1/2+tr(\bar{\rho}_{\sigma^{\rm w}_{\phi'} B}^2)-1/2
    -\cos(\phi'-\phi)tr_B[tr_A(\bar{\rho}_{AB}\sigma^{\rm w}_\phi\otimes\mathbbm{1}_B)tr_A(\bar{\rho}_{AB}\sigma_{\phi'}^{\rm w}\otimes\mathbbm{1}_B)]\nonumber\\
    =&\sin^2(\phi'-\phi)[tr(\bar{\rho}_{AB}^2)-tr(\bar{\rho}_{\sigma^{\rm p} B}^2)].
    \label{wp}
\end{align}

Let $\vec{\rho}_{\phi B}=\bar{\rho}_{\phi B}\ \vec{\phi}=tr_A(\bar{\rho}_{AB}\sigma_\phi^{\rm w}\otimes\mathbbm{1}_B)\ \vec{\phi}$, apprently $tr(\vec{\rho}_{\phi B}^{\ 2})=tr(\bar{\rho}_{\phi B}^{\ 2})=2tr(\bar{\rho}_{\sigma_\phi B}^{\ 2})-1=2I(\sigma_\phi^{\rm w}|B)$.
 \myref{wp} thus completes the proof of \myref{condwavepart}.

\section{Proof of Theorem 3}\label{appthm3}

	This proof is inspired by the works \cite{chen2014,rastegin2015} on entanglement detection with MUMs \cite{kalev2014mum}. By definition any  bipartite separable state can be written as a linear combination of product states in the form $\rho_{AB}=\sum_kp_k\rho_{A_k}\otimes\rho_{B_k}$ $(p_k>0,\ \sum_kp_k=1)$.   For a product state $\rho_A\otimes\rho_B$,  obviously  $tr(J_{i|\theta}\rho_{A}\otimes\rho_B)=\big[p^A_{i|\theta}-\frac{1}{d_A}tr(M^A_{i|\theta})\big]\cdot\big[p^B_{i|\theta}-\frac{1}{d_B}tr(M^B_{i|\theta})\big]$.  Then we have
	\begin{align}
		&J(\rho_{A}\otimes\rho_B)
		=\sum_{i,\theta}\sqrt{w_\theta}\left|p^A_{i|\theta}-tr(M^A_{i|\theta})\frac{1}{d_A}\right|\cdot\sqrt{w_\theta}\left|p^B_{i|\theta}-tr(M^B_{i|\theta})\frac{1}{d_B}\right|    \nonumber \\
		\leq&\Big[\sum_{i,\theta}w_\theta\big(p^A_{i|\theta}-tr(M^A_{i|\theta})/d_A\big)^2\Big]^{1/2} \times\Big[\sum_{i,\theta}w_\theta\big(p^B_{i|\theta}-tr(M^B_{i|\theta})/d_B\big)^2\Big]^{1/2} \nonumber\\
		\leq&\sqrt{\lVert\hat{g}_A\rVert I_{\rm com}(\rho_A)\cdot\lVert\hat{g}_B\rVert I_{\rm com}(\rho_B)} \leq\sqrt{L_A\cdot L_B},\nonumber
	\end{align}
with $L_A=\lVert\hat{g}_A\rVert(1-1/d_A)$ and $L_B=\lVert\hat{g}_B\rVert(1-1/d_B)$ being state-independent upper bounds on local information gains, and the first inequality above exploits the  Cauchy–Schwarz inequality.   Therefore, for bipartite separable states there must be  $J(\rho_{AB})=\sum_{i,\theta} w_\theta|tr(J_{i|\theta}\cdot\sum_kp_k\rho_{A_k}\otimes\rho_{B_k})|\leq \sum_kp_k\sum_{i,\theta} w_\theta|tr(J_{i|\theta}\cdot\rho_{A_k}\otimes\rho_{B_k})|=\sum_k p_kJ(\rho_{A_k}\otimes\rho_{B_k})\leq\sum_kp_k\sqrt{L_A\cdot L_B}=\sqrt{L_A\cdot L_B}$.

\section{Proof of Theorem 4}\label{appthm4}

Observe that in the case of rank-1 projective measurements \myref{gammaAB} becomes
\begin{align}
  \lVert \hat{g}\rVert+(1-\lVert \hat{g}\rVert)F^{\rm pg}(A|B)\geq \sum_{\theta}w_\theta \cdot tr(\bar{\rho}_{\mc{M}_\theta B}^2).
\end{align}
Considering that $H_{\rm min}^\varepsilon(\mc{M}_\theta|B)\geq-\log [tr(\bar{\rho}_{\mc{M}_\theta B}^2)]-\log\frac{2}{\varepsilon^2}$ (see Lemma 19 of \mycite{dupuis2014entanglement} and Theorem 7 of \mycite{tomamichel2009fully}), immediately
\begin{align}
	&\sum_\theta w_\theta H_{\rm min}^\varepsilon(\mc{M}_\theta|B)
	\geq - \log \Big[\sum_\theta w_\theta \ tr(\bar{\rho}_{\mc{M}_\theta B}^2\Big]-\log\frac{2}{\varepsilon^2}
	\geq-\log\big[ \lVert \hat{g}\rVert+(1-\lVert \hat{g}\rVert)F^{\rm pg}(A|B)\big]-\log\frac{2}{\varepsilon^2}.
\end{align}
\end{widetext}

%


\end{document}